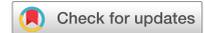

ARTICLE    OPEN

# Intriguing magnetoelectric effect in two-dimensional ferromagnetic/perovskite oxide ferroelectric heterostructure

Ping Li[1,2], Xue-Song Zhou[1] and Zhi-Xin Guo[1,3] ✉

Two-dimensional (2D) magnets have broad application prospects in the spintronics, but how to effectively control them with a small electric field is still an issue. Here we propose that 2D magnets can be efficiently controlled in a multiferroic heterostructure composed of 2D magnetic material and perovskite oxide ferroelectric (POF) whose dielectric polarization is easily flipped under a small electric field. We illustrate the feasibility of such strategy in the bilayer $CrI_3$/$BiFeO_3$(001) heterostructure by using the first-principles calculations. Different from the traditional POF multiferroic heterostructures which have strong interface interactions, we find that the interface interaction between $CrI_3$ and $BiFeO_3$(001) is van der Waals type. Whereas, the heterostructure has particular strong magnetoelectric coupling where the bilayer $CrI_3$ can be efficiently switched between ferromagnetic and antiferromagnetic types by the polarized states P↑ and P↓ of $BiFeO_3$(001). We also discover the competing effect between electron doping and the additional electric field on the interlayer exchange coupling interaction of $CrI_3$, which is responsible to the magnetic phase transition. Our results provide a avenue for the tuning of 2D magnets with a small electric field.

npj Computational Materials (2022)8:20 ; https://doi.org/10.1038/s41524-022-00706-w

## INTRODUCTION

In the past two decades, the multiferroic heterostructure composed of traditional magnetic material (such as iron, cobalt, and their alloys) and perovskite oxide ferroelectric (POF) has been widely studied[1,2], due to its great potential in realizing large magnetoelectric coupling at room temperature. However, the large number of dangling bonds at the interface easily induces significant orbital hybridizations and even the ion migration between the magnetic material and the ferroelectric oxide (strong interface interaction)[3,4]. Such strong interface interaction usually leads to the irreversible destruction of the interface magnetic structure, and ultimately reduces the cycle life of the spintronic devices[5–9].

Recently, the van de Waals (vdW) heterostructures engineering, via the stacking of layered systems with different properties, has provided a way to realize intriguing physical properties[10–14]. For example, group-IV monolayers[15–17], 1T' transition metal dichalcogenides[18], and transition metal halides/oxides[19–22], had been reported to be topological materials. Simultaneously, SnTe[23], $In_2Se_3$[24], and $CuInP_2S_6$[25] had been experimentally demonstrated to be two-dimensional ferroelectric (2DFE) materials. Meanwhile, two-dimensional ferromagnetic (2DFM) materials, such as $CrI_3$[26], $Cr_2Ge_2Te_3$[27], $Fe_3GeTe_2$[28], and $VSe_2$[29] had been successfully fabricated. Constructing heterostructures of 2DFE and 2DFM potentially provides a generally applicable route to create 2D multiferroics and magnetoelectronic couplings. Such heterostructure is expected to have vdW interface interaction due to the lack of dangling bonds, which is particular suitable for the infinite cycle life spintronic devices. While, a fundamental question is whether the vdW interlayer can induce the strong magnetoelectric coupling.

Up to now, several theoretical investigations have been performed to realize the magnetic phase transition (MPT) by manually changing the direction of dielectric polarization in the 2DFM/2DFE heterostructures, e.g., $Cr_2Ge_2Te_6$/$In_2Se_3$[30], $CrI_3$/$Sc_2CO_2$[31,32], and $MnCl_3$/$CuInP_2S_6$[33]. It had been verified that the magnetic ground state transition can be achieved even in the framework of vdW interface interaction[30–33], because the electronic structure of the 2DFM is very sensitive to the charge transfer or electric field from the 2DFE[31]. However, a much larger external electric field is required to realize MPT in the 2DFM/2DFE heterostructure, due to the nature of weak ferroelectricity of 2DFE compared to POF materials.

Here by means of density functional theory (DFT) calculations (Methods are shown in Supplementary Materials), we propose a strategy for a small electric field controlling of MPT in the 2D magnets, i.e., 2DFM/POF heterostructures. We illustrate the feasibility of such strategy in the bilayer $CrI_3$/$BiFeO_3$(001) heterostructure, where $BiFeO_3$ has much stronger electrical polarization to the 2DFM which can be flipped under a small electric field. Interestingly, we find the interface interaction between $CrI_3$ and $BiFeO_3$ is also vdW type, and the interlayer magnetic coupling of bilayer $CrI_3$ can be efficiently switched between FM and antiferromagnetic (AFM) types by flipping the dielectric polarization. Additionally, we reveal the competing effect on the MPT between electron doping and the electric field induced by the $BiFeO_3$.

## RESULTS

### $CrI_3$/$BiFeO_3$ multiferroic heterostructures

We chose bilayer $CrI_3$ with the monoclinic lattice and C2/m space group symmetry (the HT phase) as the 2DFM material[34], which exhibits the FM intralayer and AFM interlayer magnetic exchange couplings that are reported by Xu's group[26]. On the other hand, the R3c $BiFeO_3$ is chosen as the POF material due to its room temperature reversible ferroelectricity and large electric polarization. The multiferroic heterostructure was constructed by stacking

---

[1]State Key Laboratory for Mechanical Behavior of Materials, Center for Spintronics and Quantum System, School of Materials Science and Engineering, Xi'an Jiaotong University, Xi'an, Shaanxi 710049, China. [2]Key Laboratory for Computational Physical Sciences (Ministry of Eduction), Fudan University, Shanghai 200433, China. [3]Key Laboratory of Polar Materials and Devices (Ministry of Education), East China Normal University, Shanghai 200241, China. ✉email: zxguo08@xjtu.edu.cn

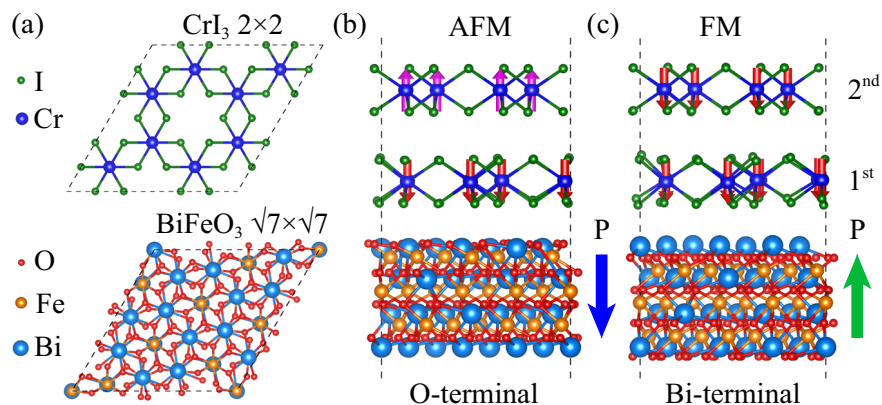

**Fig. 1 The magnetic phase transition was induced by polarization reversal. a** Top views of the 2 × 2 CrI$_3$ and $\sqrt{7} \times \sqrt{7}$ BiFeO$_3$(001). **b, c** Side views of CrI$_3$/BiFeO$_3$ heterostructures with the top layer O/Bi atom of BiFeO$_3$ (denoted as O-terminal and Bi-terminal), respectively. The red and magenta arrows in bilayer CrI$_3$ denote the spin directions. The blue and green arrows denotes the direction of the out-of-plane electric polarization in BiFeO$_3$(001), respectively.

**Table 1.** Calculated total energies of CrI$_3$/BiFeO$_3$(001) heterostructure with CrI$_3$ in FM ($E_{FM}$) and AFM ($E_{AFM}$) interlayer-coupling phases.

| | | Bridge | Hollow | Top |
|---|---|---|---|---|
| Bi-terminal | $E_{FM}$ (meV) | 25.49 | 0.00 | 14.18 |
| | $E_{AFM}$ (meV) | 35.66 | 20.56 | 28.30 |
| | $\Delta E$ (meV) | −10.17 | −20.56 | −14.12 |
| | $E_b$ (eV) | 0.41 | 0.34 | 0.34 |
| | $l$ (Å) | 3.75 | 3.79 | 3.76 |
| O-terminal | $E_{FM}$ (meV) | 9.17 | 833.09 | 870.44 |
| | $E_{AFM}$ (meV) | 0.00 | 822.49 | 858.40 |
| | $\Delta E$ (meV) | 9.17 | 10.60 | 12.04 |
| | $E_b$ (eV) | 0.51 | 0.36 | 0.37 |
| | $l$ (Å) | 3.18 | 3.30 | 3.31 |

The total energies are defined relative to that of FM phase in hollow configuration for CrI$_3$ on Bi-terminal BiFeO$_3$(001) and AFM phase in bridge configuration for CrI$_3$ on O-terminal BiFeO$_3$(001), respectively. $\Delta E$ is defined as $E_{FM} - E_{AFM}$, and $E_b$ is the binding energy of the ground state of CrI$_3$/BiFeO$_3$ heterostructure. $l$ represents the average bond length between Bi and I atoms of the interface for Bi-terminal and O and I atoms of the interface for O-terminal.

bilayer CrI$_3$ on BiFeO$_3$(001) surface. Considering that the optimized lattice constant of CrI$_3$ is 7.00 Å and that of BiFeO$_3$(001) is 5.64 Å, the 2 × 2 unit cell of CrI$_3$ is commensurate to the $\sqrt{7} \times \sqrt{7}$ surface unit cell of BiFeO$_3$(001) with a lattice mismatch of 6.59%, as shown in Fig. 1(a). We have considered three typical stacking configurations for CrI$_3$ on O(Bi)-terminal of BiFeO$_3$(001), i.e., the bottom CrI$_3$ layer being on top, bridge, and hollow positions of the top O(Bi)-terminal layer (see Supplementary Fig. 1). After fully structural optimization, we find that the hollow (bridge) structure is the most stable for Bi-terminal (O-terminal) heterostructure (see Table 1).

To quantify the interaction between CrI$_3$ and BiFeO$_3$(001), we calculated the binding energy $E_b$[35], defined as $E_b = (E_{CrI3} + E_{BiFeO3} - E_{tot})/N_I$. Here $E_{tot}$, $E_{CrI3}$, and $E_{BiFeO3}$ are the total energies of the CrI$_3$/BiFeO$_3$(001), the bilayer CrI$_3$, and the clean BiFeO$_3$(001) surface, respectively. $N_I$ is the number of I atoms at the interface. The calculated $E_b$ are shown in Table 1. For all the structures, the binding energies (0.34–0.51 eV I$^{-1}$) are comparable to that of CrI$_3$ on semiconductor substrate (0.28–0.41 eV Cr$^{-1}$), implying the nature of vdW interaction between CrI$_3$ and BiFeO$_3$[14]. Moreover, the average Bi–I (O–I) bond lengths at the interface are in the range of 3.75–3.79 (3.18–3.31) Å for Bi-terminal (O-terminal).

Considering that the atomic radii of I, O and Bi atoms are 1.40 Å, 0.66 Å, and 1.56 Å, respectively, this result confirms the non-bonded interface interactions. We have additionally calculated charge density distributions of the CrI$_3$/BiFeO$_3$(001) heterostructure (Supplementary Fig. 2) and found there is little charge density overlap at the interface. Therefore, one can conclude that the interface interaction between CrI$_3$ and BiFeO$_3$ is vdW type, where the migration of ions from BiFeO$_3$ to CrI$_3$ during the ferroelectric polarization inversion hardly occurs. Such multiferroic heterostructure is expected to be free of the flipping cycle-life issue.

### Tunable interlayer coupling of bilayer CrI$_3$ and mechanism for interfacial multiferroicity

The total energy calculations further show that BiFeO$_3$(001) can significantly affect the magnetic interlayer coupling of bilayer CrI$_3$. As shown in Fig. 1, both the FM and AFM interlayer magnetic configurations of bilayer CrI$_3$ had been considered. It is found that the total energies of their obviously depend on the dielectric polarized directions of BiFeO$_3$(001), i.e., P↓ polarization when the O atoms move up along [001] direction (O-terminal, Fig. 1b), and P↑ polarization on the contrary (Bi-terminal, Fig. 1c). As shown in Table 1, among all the stacking configurations (bridge/hollow/top) the FM (AFM) interlayer-coupling phase of bilayer CrI$_3$ always has the lowest total energy when BiFeO$_3$(001) is in the P↑ (P↓) state. This feature shows that the magnetic phase of bilayer CrI$_3$ can be well switched by flipping the dielectric polarized directions of BiFeO$_3$(001), which is easily realized under a small electric field. Note that the energy differences $\Delta E$ between FM and AFM states in both the P↑ (P↓) states are in range of 9.2–20.6 meV. The distinguish energy differences of FM and AFM states with ferroelectric polarization flipping indicate that the system has a strong magnetoelectric coupling.

It is desirable to understand the origin of MPT from AFM to FM for CrI$_3$ on BiFeO$_3$(001). In general, three combined factors induced by the BiFeO$_3$ substrate can be responsible to the MPT: (1) structure reconstruction (deformation) of CrI$_3$ induced by BiFeO$_3$, (2) charge transfer effect from BiFeO$_3$, and (3) electric field (electric polarization) effect from BiFeO$_3$. In the following, we will distinguish the roles of these factors on the MPT via studying the bilayer CrI$_3$ detached from BiFeO$_3$. The DFT calculations show that the vdW interaction from BiFeO$_3$ gives rise to the structure reconstruction of bottom CrI$_3$ layer to a certain extent (see Fig. 1c). In order to clarify the influence of structure reconstruction to interlayer magnetic coupling, we calculated the total energy differences between FM and AFM phases of bilayer CrI$_3$ detached from $BiFeO_3(001)$. As shown in Table 2, the energy differences ($\Delta E = E_{FM} - E_{AFM}$) of all the $CrI_3$ in different configurations are positive. This feature means that the MPT in bilayer $CrI_3$ is not attributed to the structure reconstruction. In addition, we have calculated the most stable phase of unrelaxed bilayer $CrI_3$ structures, and found it has the same magnetic ground state, i.e., AFM state, as that of the relaxed one (see Supplementary Table 1). This result confirms that the geometry distortion of bilayer $CrI_3$ induced by surface as well as other effects do not change the magnetic ground state.

Table 2. Calculated total energies of $CrI_3$ detached from $BiFeO_3(001)$ in FM ($E_{FM}$) and AFM ($E_{AFM}$) interlayer-coupling phases, which is defined relative to that of AFM phase in bridge (hollow) configuration for $CrI_3$ detached from Bi-terminal (O-terminal) $BiFeO_3(001)$.

|  |  | Bridge | Hollow | Top |
|---|---|---|---|---|
| Bi-terminal | $E_{FM}$ (meV) | 12.27 | 96.40 | 17.38 |
|  | $E_{AFM}$ (meV) | 0.00 | 78.00 | 3.69 |
|  | $\Delta E$ (meV) | 12.27 | 18.40 | 13.69 |
| O-terminal | $E_{FM}$ (meV) | 316.78 | 17.17 | 38.44 |
|  | $E_{AFM}$ (meV) | 300.23 | 0.00 | 21.32 |
|  | $\Delta E$ (meV) | 16.55 | 17.17 | 17.12 |

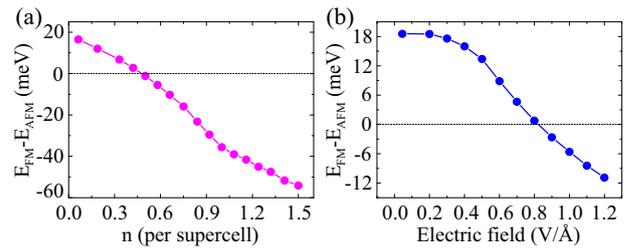

Fig. 3 Electric field and electron doping tune the magnetic phase transition of bilayer $CrI_3$. Energy differences between AFM and FM phases for bilayer $CrI_3$ in hollow configuration detached from Bi-terminal $BiFeO_3(001)$ under (a) electron doping and (b) electric field.

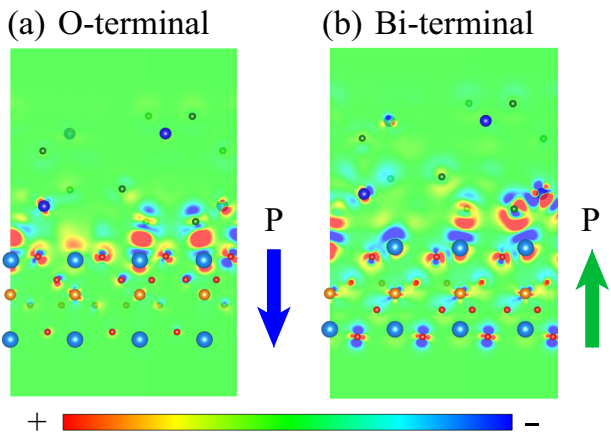

Fig. 2 2D charge density difference of $CrI_3/BiFeO_3(001)$ heterostructure in hollow configuration. a O-terminal (P↓ polarization) and b Bi-terminal (P↑ polarization), respectively. The red and blue areas indicate gain and loss electrons, respectively. The isosurface value of charge density difference is 0.001 e Bohr$^{-3}$.

Then, we explored the charge transfer effect on the MPT of bilayer $CrI_3$. Figure 2a, b shows the calculated differential charge density distributions of $CrI_3/BiFeO_3$ heterostructure, where obviously charge transfer from $CrI_3$ ($BiFeO_3$) to $BiFeO_3$ ($CrI_3$) appears for P↓ (P↑) polarization. We have additionally estimated the amount of transferred charge and found it is sizable in both cases, i.e., 0.48 e per supercell from $CrI_3$ to $BiFeO_3$ for P↓ and 1.63 e per supercell from $BiFeO_3$ to $CrI_3$ for P↑, respectively. In order to understand the mechanism of the charge transfer, we calculated the density of states (DOS) by using the Heyd–Scuseria–Ernzerhof (HSE) method sketched in Supplementary Fig. 3. For the O-terminal case, there are unsaturated O atoms in the top layer of $BiFeO_3$ at the interface, which makes the surface state of $BiFeO_3$ becomes metallic [see Supplementary Figure 3(b)]. Therefore, the charge transfer at the interface is mainly determined by the electronegativity of the interface atoms. The electronegativity of O (3.44) atom is larger than that of I (2.66) atom, which leads to the 0.48 e per supercell transferred from $CrI_3$ to $BiFeO_3(001)$. For the Bi-terminal case, on the other hand, the O atoms on the top layer are saturated by Bi atoms (see Fig. 1), which induces the band gap for the interface state and thus allows the HSE investigations. As shown in Supplementary Figure 3(f), the valence-band maximum of $BiFeO_3(001)$ overlaps with the conduction-band minimum (CBM) of $CrI_3$, which corresponds to a type III band alignment. Such phenomenon is attributed to the electron transfer from $BiFeO_3$ to $CrI_3$, which results in the upshift (downshift) of CBM energy level of $BiFeO_3$ ($CrI_3$) in the heterostructure in comparation with the isolated ones [see Supplementary Fig. 3(d–f)].

To further clarify the mechanism of electron doping on MPT of bilayer $CrI_3$, we have systematically calculated the electron doping concentration (n) dependence of total energy differences between FM and AFM phases ($\Delta E$) for bilayer $CrI_3$ detached from Bi-terminal $BiFeO_3(001)$ with hollow configuration. Here we simulated the electron-doping effect through doping H atom in the hollow position of the first $CrI_3$ (see Supplementary Fig. 4). As shown in Fig. 3a, $\Delta E$ monotonically decreases with doping concentration increases, where the MPT from AFM to FM appears when n > 0.5 e per supercell. This value is much smaller than that transferred from $BiFeO_3$ to $CrI_3$ (1.63 e per supercell), thus the charge transfer is expected to be a dominating factor to the MPT of bilayer $CrI_3$.

In addition, we have explored the effect of electric field on MPT of bilayer $CrI_3$, which is originated from the P↑ polarization of $BiFeO_3(001)$. As shown in Fig. 3b, $\Delta E$ of detached $CrI_3$ monotonically decreases with the increase of electric field, where the MPT occurs when the electric field becomes larger than 0.9 V Å$^{-1}$. On the other hand, the equivalent electric field from $BiFeO_3(001)$ is estimated to be about 0.15 V Å$^{-1}$. This result shows that the electric field from $BiFeO_3$ has minor effect on the MPT of bilayer $CrI_3$.

To further clarify the combined effect of charge transfer and the additional electric field on the MPT of bilayer $CrI_3$, we have investigated the total energy difference $\Delta E$ of bilayer $CrI_3$ detached from $BiFeO_3(001)$ in three cases, i.e., Case 1 without electric field and electron doping (only structure reconstruction effect is considered), Case 2 with only electron doping (1.5 e), and Case 3 with electron doping (1.5 e) and the additional electric field (0.5 V Å$^{-1}$). As shown in Table 3, $\Delta E$ significantly decreases from 18.40 meV (Case 1) to −54.09 meV (Case 2) with only electron doping, whereas it increases to −3.63 meV (Case 3) with the additional electric field 0.5 V Å$^{-1}$. This result shows that the combined effect of electron doping and electric field on the MPT of bilayer $CrI_3$ is inferior to that of the electron doping alone, although the electric field alone also benefits to the MPT from AFM to FM phase. This feature shows that the electric field has an effect of weakening the FM phase in an electron-doping environment.

To reveal the underlying mechanism of charge transfer and electric field on the MPT of bilayer $CrI_3$, we explored the interlayer exchange interactions between Cr atoms with both FM and AFM spin configurations. As indicated in Fig. 4a, the hopping between $t_{2g}$ and $t_{2g}$ orbitals is allowed for the AFM spin configuration but prohibited for the FM spin configuration according to the Pauli

**Table 3.** Calculated Heisenberg exchange parameters (meV) of bilayer CrI$_3$ detached from Bi-terminal BiFeO$_3$(001) (hollow configuration).

|        | $J_{1\|\|}$ | $J_{2\|\|}$ | $J_{\perp 1}$ | $J_{\perp 2}$ | $\bar{J}_\perp$ | $\Delta E$ |
|--------|-------------|-------------|---------------|---------------|-----------------|-----------|
| Case 1 | −4.38       | −3.91       | 1.35          | −1.09         | 0.26            | 18.40     |
| Case 2 | −4.94       | −4.53       | 1.32          | −2.08         | −0.76           | −54.09    |
| Case 3 | −5.37       | −4.13       | 1.17          | −1.22         | −0.05           | −3.63     |

Case 1, Case 2, and Case 3 correspond to the parameters obtained without electric field and electron doping, with only electron doping (1.5 e), and with both electric field (0.5 V Å$^{-1}$) and electron doping (1.5 e), respectively. $J_{1\|\|}$ ($J_{2\|\|}$) represents the Heisenberg exchange parameter for the intralayer interaction between Cr atoms in the first (second) layer. $J_{\perp 1}$ and $J_{\perp 2}$ are the nearest-neighboring and second-neighboring interlayer exchange parameters between Cr atoms, respectively. $\bar{J}_\perp$ is defined as $J_{\perp 1} + J_{\perp 2}$.

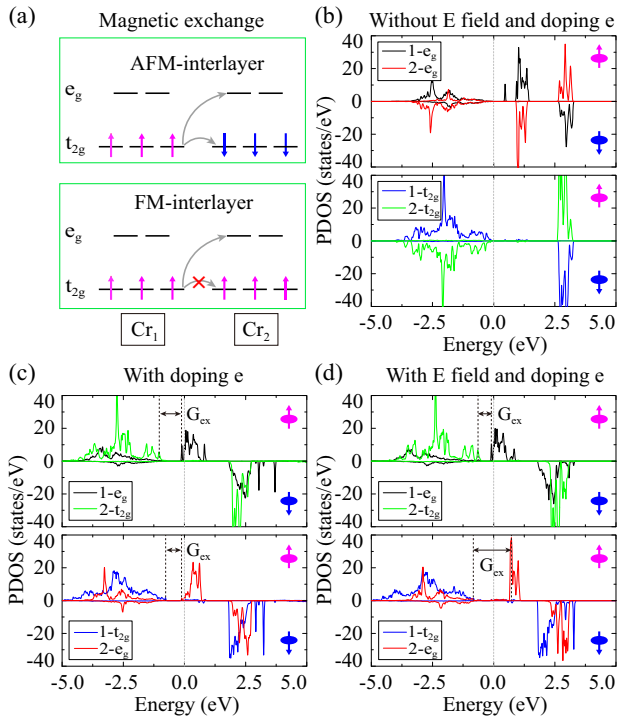

**Fig. 4 Interlayer spin-exchange model and calculated PDOS for bilayer CrI$_3$ detached from Bi-terminal BiFeO$_3$(001) (hollow configuration). a** Schematic illustration of the orbital-dependent interlayer interactions. The hopping form $t_{2g}$–$t_{2g}$ is allowed in the AFM exchange, whereas it is prohibited in the FM exchange. **b**–**d** PDOS for $e_g$ and $t_{2g}$ orbitals of Cr in bilayer CrI$_3$. **b** PDOS of AFM phase without electric field and electron doping (Case 1), **c** PDOS of FM phase with electron doping (Case 2), and **d** PDOS of FM phase with electron doping and the additional electric field (Case 3). The 1-$e_g$ and 1-$t_{2g}$ (2-$e_g$ and 2-$t_{2g}$) denote the $e_g$ and $t_{2g}$ orbitals of Cr in the first layer (second layer), respectively. The exchange interaction strength between $e_g$ and $t_{2g}$ orbitals is inversely proportional to the virtual exchange gap $G_{ex}$. When the electric field is additionally applied to the electron-doped CrI$_3$, $G_{ex}$ between 2-$e_g$ and 1-$t_{2g}$ significantly increases.

exclusion principle. This feature means that the $t_{2g}$–$t_{2g}$ ($t_{2g}$–$e_g$) orbital hybridizations give rise to the AFM (FM) phase of bilayer CrI$_3$ from the viewpoint of the Hund's coupling[36,37]. Figure 4(b–d) further shows the calculated the projected density of states (PDOS) of bilayer CrI$_3$ in Cases 1–3 discussed above, where the 1-$e_g$ and 1-$t_{2g}$ (2-$e_g$ and 2-$t_{2g}$) represent first layer (second layer) Cr $e_g$ and $t_{2g}$ orbitals, respectively. As one can see from Fig. 4b, the interlayer $t_{2g}$-$t_{2g}$ orbitals have strong hybridization without electron doping and electric field (Case 1), in agreement with the AFM spin-exchange model in the upper panel of Fig. 4a. When only electron doping is applied to bilayer CrI$_3$ (Case 2), the $t_{2g}$–$t_{2g}$ orbital hybridization becomes much weaker, whereas the $t_{2g}$–$e_g$ hybridization becomes obviously stronger due to the downshift of $e_g$ energy levels (the exchange interaction strength between $e_g$ and $t_{2g}$ orbitals is inversely proportional to the virtual exchange gap ($G_{ex}$), Fig. 4c). This characteristic is consistent with the FM spin-exchange model in the under panel of Fig. 4a. When the electric field is additionally applied to the electron-doped CrI$_3$ (Case 3), $G_{ex}$ significantly increases (under panel of Fig. 4d), indicating that the $t_{2g}$–$e_g$ orbitals hybridization becomes weaker. This feature explains why the electric field in the electron doping environment can weaken the FM phase discussed above.

To quantitatively investigate the effects of electron doping and the additional electric field on the interlayer exchange coupling of bilayer CrI$_3$, we constructed a Heisenberg model for the bilayer CrI$_3$ detached from Bi-terminal BiFeO$_3$(001) (hollow configuration) with the Hamiltonian as

$$H = E_0 + \sum_{i,j} J_{1\|\|} S_i \cdot S_j + \sum_{k,l} J_{2\|\|} S_k \cdot S_l + \sum_{i,k} J_\perp S_i \cdot S_k, \quad (1)$$

where $E_0$ is the ground state energy independent of the spin configurations. $S_i$, $S_j$, $S_k$, and $S_l$ represent the magnetic moments at sites $i$, $j$, $k$, and $l$, respectively. $J_{\|\|}$ and $J_\perp$ denote the intralayer and interlayer exchange interactions, respectively. We adopted $J_{1\|\|}$ and $J_{2\|\|}$ to denote the nearest-neighbor intralayer Cr-Cr exchange interaction of the first and second CrI$_3$ layer, respectively. $J_{\perp 1}$ and $J_{\perp 2}$ are the nearest-neighbor and second-neighbor interlayer Cr-Cr exchange interactions [Supplementary Fig. 5(a)], respectively. The details for the calculation of exchange parameters based on the above Hamiltonian and total energy calculations using DFT are shown in the Supplementary material (see Supplementary Fig. 5 and Supplementary Table 2).

The obtained results for Cases 1–3 are summarized in Table 3. As one can seen, in all cases $J_{\perp 1} > 0$ (contributes to AFM phase) and $J_{\perp 2} < 0$ (contributes to FM phase). This feature shows that $J_{\perp 1}$ is dominated by exchange interaction between Cr half-filled $t_{2g}$ orbitals which induces an AFM coupling, while $J_{\perp 2}$ is dominated by a virtual excitation between the Cr half-filled $t_{2g}$ orbitals and the empty $e_g$ orbitals which leads to a FM coupling. Therefore, the ground state of bilayer CrI$_3$ is determined by the competition between nearest-neighbor and second-neighbor interlayer Cr–Cr exchange interactions. Considering that the number of nearest-neighbor and second-neighbor interlayer Cr–Cr interaction pairs are the same all the three cases, the MPT of CrI$_3$ can be predicted by the value of $\bar{J} = J_{\perp 1} + J_{\perp 2}$. In case without electric field and electron doping (Case 1), $\bar{J} = 0.26$ meV, indicating the AFM ground state from the Heisenberg model which agrees well with the direct DFT calculations[38]. When the electron doping is applied (Case 2), the $J_{\perp 1}$ decreases little whereas $J_{\perp 2}$ nearly has a double enhancement (-1.09 to -2.08 meV), leading to a strong FM ground state ($\bar{J} = -0.75$ meV). Additionally, when both the electron doping and the electric field are applied (Case 3), $J_{\perp 2}$ significantly changes from −2.08 to −1.22 meV but $J_{\perp 1}$ only decreases a little, resulting a weak FM ground state ($\bar{J} = -0.05$ meV). These results reveal that electron doping and the additional applied electric field have significant but opposite influence on the second-neighbor interlayer Cr–Cr exchange interaction, the variation of which is the origin of MPT in bilayer CrI$_3$. We have also calculated the exchange parameters in the three cases for the bilayer CrI$_3$ without geometry reconstruction, and obtained the similar results (see Supplementary Table 3 and 4), which confirms the intrinsic nature of competition between electron doping and the additional electric field on the interlayer exchange coupling of bilayer CrI$_3$.

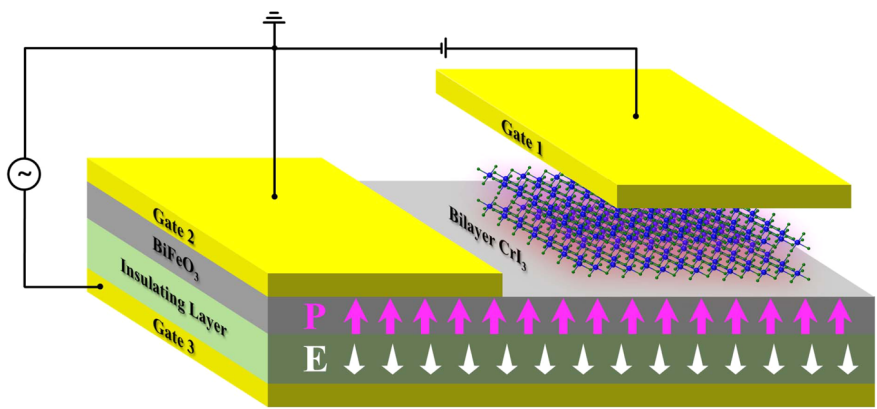

**Fig. 5 Schematic image of spin valve device based on 2DFM (CrI$_3$)/POF (BiFeO$_3$) heterostructure.** Gate 1 and Gate 2 are the bias electrodes to read the electronic resistance, Gate 3 is the gating electrode to control electron polarization of POF.

We have also explored the influence of spin-orbit coupling (SOC) on the magnetic ground state of CrI$_3$ in the heterostructure, where the hollow configuration under different electrical polarizations was considered. The calculated results are summarized in Supplementary Table 5. It is seen that the FM state has the lowest total energy in the Bi-terminal configuration, and the AFM interlayer coupling becomes the ground state in the O-terminal configuration. This result is qualitatively consistent with that obtained without SOC, implying that SOC effect does not change the conclusion in our investigation.

Additionally, considering that there is also possibility for CrI$_3$ laying on Fe-terminal BiFeO$_3$[39,40], we have further explored the switching of magnetic configurations in such heterostructure (Supplementary Figure 6). As shown in Supplementary Table 6, the FM (AFM) interlayer-coupling phase of bilayer CrI$_3$ has lower total energy when the BiFeO$_3$(001) is in the P↑ (P↓) state, showing the effective switching of magnetic states with ferroelectric polarization. This result is consistent with that of CrI$_3$ on O/Bi-terminal BiFeO$_3$(001) discussed above.

## DISCUSSION
Before closing, we discuss the potential applications of 2DFM/POF multiferroic heterostructure in the non-volatile spintronic devices, i.e., spin value. The traditional spin valve composes of two magnetic materials separated by a nonmagnetic spacer, where the low electronic resistance ($R_1$) appears when their spin directions are parallel, whereas the high electronic resistance ($R_2$) appears when their spin directions are antiparallel. The resistance change is a result of the giant magnetoresistance[41] or tunnel magnetoresistance[42] effect. The relative spin directions of the two magnetic materials are generally controlled by external magnetic field or spin torques (STT[43], SOT[44]) induced by the electron current in the chip, which significantly limits the performance of spin values owning to the Joule heating effect. Here we propose that an electron-current free spin valve can be achieved by using the 2DFM (CrI$_3$)/POF (BiFeO$_3$) heterostructure, where the nonmagnetic spacer is the vacuum layer induced by the vdW interaction between two magnetic layers of 2DFM (Fig. 5). As shown in Fig. 5, Gate 1 and Gate 2 are the bias electrodes to read the electronic resistance, which is mainly contributed by $R_1$ and $R_2$. Gate 3 is the gating electrode to control electron polarization of POF with a small electric field, which in turn effectively controls the relative spin directions of 2DFM and thus the electronic resistance. Note that the switching between $R_1$ or $R_2$) is achieved only by flipping the direction of electric field in Gate 3, where little charge current and Joule heat is produced. In addition to the room temperature ferroelectricity of POF, a high-performance non-volatile spin value composed of 2DFM/POF heterostructure is likely to be achieved.

## METHODS
First, we used the Device Studio to build structure and the RESCU to roughly optimize the structure[45]. Then, we implemented the Vienna *Abinitio* Simulation Package[46,47] for the first-principles calculations based on DFT. The electron exchange-correlation functional was described by the generalized gradient approximation of the Perdew-Burke-Ernzerhof functional[48]. The vdW density functional (optB86b-vdW) which was capable of treating the dispersion force, was adopted for the exchange-correlation functional[49,50]. The plane-wave basis set with a kinetic energy cutoff of 450 eV was employed. The $3 \times 3 \times 1\Gamma$-centered $k$ meshes were adopted for the structural optimization. The geometry optimization was performed until the remaining Hellmann–Feynman forces become less than 0.01 eV/Å to obtain the final structures. The electronic structures and magnetic properties were calculated using $6 \times 6 \times 1\Gamma$-centered $k$ meshes. Moreover, to describe the strongly correlated 3$d$ orbitals of the Fe and Cr atoms, the GGA+U method was used[51]. The onsite Coulomb interaction was considered for Fe and Cr 3$d$ orbitals by setting effective Hubbard U to 3.8 eV and 2 eV, respectively. The BiFeO$_3$ surface was simulated by a repeating slab model consisting of a four-BiFeO$_3$-layer slab cleaved from the $R3c$ BiFeO$_3$(001). The BiFeO$_3$ slab was separated from its images by the 20 Å vacuum layer with bilayer CrI$_3$ placed on top. The CrI$_3$ with $2 \times 2$ periodicity were deposited on the $\sqrt{7} \times \sqrt{7}$ BFeO$_3$(001) surface in the simulation. Moreover, the linear alinement of spin polarization of Fe atoms in BiFeO$_3$ had been considered, where the same initial magnetic momentum was adopted for all the Fe atoms in the same layer and adjacent layers are oppositely magnetized.

In addition, we simulated the electron doping effect through doping H atoms in the hollow position of the first CrI$_3$ layer, where the bilayer CrI$_3$ was attached from Bi-terminal BiFeO$_3$ (hollow configuration). The electron doping concentration was controlled via changing the number of H atoms and the number of electron per H atom in the bilayer CrI$_3$ supercell. Supplementary Figure 4 further illustrates the structure of bilayer CrI$_3$ with 2 H atoms doped and the number of electron per H atom is 0.75 e, which corresponds to a 1.5 e per supercell doping concentration in the main text.

## DATA AVAILABILITY
The data that support the findings of this study are available from the corresponding author upon reasonable request.

## CODE AVAILABILITY
The codes are available from the corresponding author upon reasonable request.

## REFERENCES
1. Heron, J. T. et al. Electric-field-induced magnetization reversal in a ferromagnet-multiferroic heterostructure. *Phys. Rev. Lett.* **107**, 217202 (2011).
2. Zhang, S. et al. Electric-field control of nonvolatile magnetization in $Co_{40}Fe_{40}B_{20}Pb(Mg_{1/3}Nb_{2/3})_{0.7}Ti_{0.3}O_{0.3}$ structure at room temperature. *Phys. Rev. Lett.* **108**, 137203 (2012).
3. Oleinik, I. I., Tsymbal, E. Y. & Pettifor, D. G. Atomic and electronic structure of Co/$SrTiO_3$/Co magnetic tunnel junctions. *Phys. Rev. B* **65**, 020401(R) (2001).
4. Duan, C. G., Jaswal, S. S. & Tsymbal, E. Y. Predicted magnetoelectric effect in Fe/$BaTiO_3$ multilayers: ferroelectric control of magnetism. *Phys. Rev. Lett.* **97**, 047201 (2006).
5. Heron, J. T. et al. Deterministic switching of ferromagnetism at room temperature using an electric field. *Nature* **516**, 370 (2014).
6. Zhang, Q. et al. Polarization-mediated thermal stability of metal/oxide heterointerface. *Adv. Mater.* **27**, 6934 (2015).
7. Yang, S. W. et al. Non-volatile 180° magnetization reversal by an electric field in multiferroic heterostructures. *Adv. Mater.* **26**, 7091 (2014).
8. Li, P. et al. Electric field manipulation of magnetization rotation and tunneling magnetoresistance of magnetic tunnel junctions at room temperature. *Adv. Mater.* **26**, 4320 (2014).
9. Waldrop, M. M. More than Moore. *Nature* **530**, 144 (2016).
10. Li, P., Guo, Z. X. & Cao, J. X. A new approach for fabricating germanene with Dirac electrons preserved: a first principles study. *J. Mater. Chem. C* **4**, 1736 (2016).
11. Li, P. et al. Topological Dirac states beyond π-orbitals for silicene on SiC(0001) surface. *Nano Lett.* **17**, 6195 (2017).
12. Gong, C. & Zhang, X. Two-dimensional magnetic crystals and emergent heterostructure devices. *Science* **363**, 706 (2019).
13. Li, P. Stanene on a SiC(0001) surface: a candidate for realizing quantum anomalous Hall effect. *Phys. Chem. Chem. Phys.* **21**, 11150 (2019).
14. Liu, N., Zhou, S. & Zhao, J. High-Curie-temperature ferromagnetism in bilayer $CrI_3$ on bulk semiconducting substrates. *Phys. Rev. Mater.* **4**, 094003 (2020).
15. Kane, C. L. & Mele, E. J. Quantum spin Hall effect in graphene. *Phys. Rev. Lett.* **95**, 226801 (2005).
16. Liu, C. C., Feng, W. X. & Yao, Y. G. Quantum spin Hall effect in silicene and two-dimensional germanium. *Phys. Rev. Lett.* **107**, 076802 (2011).
17. Xu, Y. et al. Large-gap quantum spin Hall insulators in tin films. *Phys. Rev. Lett.* **111**, 136804 (2013).
18. Qian, X., Liu, J., Fu, L. & Li, J. Quantum spin Hall effect in two-dimensional transition metal dichalcogenides. *Science* **346**, 1344 (2014).
19. Zhou, L. et al. New family of quantum spin Hall insulators in two-dimensional transition-metal halide with large nontrivial band gaps. *Nano Lett.* **15**, 7867 (2015).
20. Li, P. Prediction of intrinsic two dimensional ferromagnetism realized quantum anomalous Hall effect. *Phys. Chem. Chem. Phys.* **21**, 6712 (2019).
21. Li, P. & Cai, T. Y. Two-dimensional transition-metal oxides $Mn_2O_3$ realized the quantum anomalous Hall effect. *J. Phys. Chem. C* **124**, 12705 (2020).
22. Li, P., Ma, Y., Zhang, Y. & Guo, Z. X. Room temperature quantum anomalous hall insulator in a Honeycomb-Kagome lattice, $Ta_2O_3$, with huge magnetic anisotropy energy. *ACS Appl. Electron. Mater.* **3**, 1826 (2021).
23. Chang, K. et al. Discovery of robust in-plane ferroelectricity in atomic-thick SnTe. *Science* **353**, 6296 (2016).
24. Zhou, Y. et al. Out-of-plane piezoelectricity and ferroelectricity in layered α-$In_2Se_3$ nanoflakes. *Nano Lett.* **17**, 5508 (2017).
25. Liu, F. et al. Room-temperature ferroelectricity in $CuInP_2S_6$ ultrathin flakes. *Nat. Commun.* **7**, 12357 (2016).
26. Huang, B. et al. Layer-dependent ferromagnetism in a van der Waals crystal down to the monolayer limit. *Nature* **546**, 270 (2017).
27. Gong, C. et al. Discovery of intrinsic ferromagnetism in two-dimensional van der Waals crystals. *Nature* **546**, 265 (2017).
28. Deng, Y. et al. Gate-tunable room-temperature ferromagnetism in two-dimensional $Fe_3GeTe_2$. *Nature* **563**, 94 (2018).
29. Bonilla, M. et al. Strong room-temperature ferromagnetism in $VSe_2$ monolayers on van der Waals substrates. *Nat. Nanotechnol.* **13**, 289 (2018).
30. Gong, C., Kim, E. M., Wang, Y., Lee, G. & Zhang, X. Multiferroicity in atomic van der Waals heterostructures. *Nat. Commun.* **10**, 2657 (2019).
31. Zhao, Y., Zhang, J. J., Yuan, S. & Chen, Z. Nonvolatile electrical control and heterointerface-induced half-metallicity of 2D ferromagnets. *Adv. Funct. Mater.* **29**, 1901420 (2019).
32. Lu, Y. et al. Artificial multiferroics and enhanced magnetoelectric effect in van der Waals heterostructures. *ACS Appl. Mater. Interfaces* **12**, 6243 (2020).
33. Li, Z. & Zhou, B. Theoretical investigation of nonvolatile electrical control behavior by ferroelectric polarization switching in two-dimensional $MnCl_3$/$CuInP_2S_6$ van der Waals heterostructures. *J. Mater. Chem. C* **8**, 4534 (2020).
34. McGuire, M. A., Dixit, H., Cooper, V. R. & Sales, B. C. Coupling of crystal structure and magnetism in the layered, ferromagnetic insulator $CrI_3$. *Chem. Mater.* **27**, 612 (2015).
35. Guo, Z. X., Furuya, S., Iwata, J. I. & Oshiyama, A. Absence and presence of Dirac electrons in silicene on substrates. *Phys. Rev. B* **87**, 235435 (2013).
36. Goodenough, J. B. Theory of the role of covalence in the perovskite-type manganites [La, M(II)]$MnO_3$. *Phys. Rev.* **100**, 564 (1955).
37. Kanamori, J. Superexchange interaction and symmetry properties of electron orbitals. *J. Phys. Chem. Solids* **10**, 87 (1959).
38. Sivadas, N., Okamoto, S., Xu, X., Fennie, C. J. & Xiao, D. Stacking-dependent magnetism in bilayer $CrI_3$. *Nano Lett.* **18**, 7658 (2018).
39. Dai, J. Q., Xu, J. W. & Zhu, J. H. Thermodynamic stability of $BiFeO_3$(0001) surface from ab Inito theory. *ACS Appl. Mater. Interfaces* **9**, 3168 (2017).
40. Dai, J. Q., Xu, J. W. & Zhu, J. H. First-principles study on the multiferroic $BiFeO_3$(0001) polar surfaces. *Appl. Surf. Sci.* **392**, 135 (2017).
41. Baibich, M. N. et al. Giant magnetoresistance of (001)Fe/(001)Cr magnetic superlattices. *Phys. Rev. Lett.* **61**, 2472 (1988).
42. Moodera, J. S., Kinder, L. R., Wong, T. M. & Meservey, R. Large magnetoresistance at room temperature in ferromagnetic thin film tunnel junctions. *Phys. Rev. Lett.* **74**, 3273 (1995).
43. Diao, Z. et al. Spin-transfer torque switching in magnetic tunnel junctions and spin-transfer torque random access memory. *J. Phys. Condens. Matter* **19**, 165209 (2007).
44. Lau, Y. C., Betto, D., Rode, K., Coey, J. M. D. & Stamenov, P. Spin-orbit torque switching without an external field using interlayer exchange coupling. *Nat. Nanotechol.* **11**, 758 (2016).
45. Michaud-Rioux, V., Zhang, L. & Guo, H. RESCU: a real space electronic structure method. *J. Comput. Phys.* **307**, 593 (2016).
46. Kresse, G. & Hafner, J. Ab initio molecular dynamics for liquid metals. *Phys. Rev. B* **47**, 558(R) (1993).
47. Kresse, G. & Joubert, D. From ultrasoft pseudopotentials to the projector augmented-wave method. *Phys. Rev. B* **59**, 1758 (1999).
48. Perdew, J. P., Burke, K. & Ernzerhof, M. Generalized gradient approximation made simple. *Phys. Rev. Lett.* **77**, 3865 (1996).
49. Dion, M., Rydberg, H., Schroder, E., Langreth, D. C. & Lundqvist, B. I. Van der Waals density functional for general geometries. *Phys. Rev. Lett.* **92**, 246401 (2014).
50. Klimes, J., Bowler, D. R. & Michaelides, A. Van der Waals density functionals applied to solids. *Phys. Rev. B* **83**, 195131 (2011).
51. Dudarev, S. L., Botton, G. A., Savrasov, S. Y., Humphreys, C. J. & Sutton, A. P. Electron-energy-loss spectra and the structural stability of nickel oxide: an LSDA +U study. *Phys. Rev. B* **57**, 1505 (1998).

## ACKNOWLEDGEMENTS
This work is supported by National Natural Science Foundation of China (Nos. 12074301 and 12004295), National Key R&D Program of China (2018YFB0407600), the Science Fund for Distinguished Young Scholars of Hunan Province (No. 2018JJ1022). P.L. thanks China's Postdoctoral Science Foundation funded project (No. 2020M673364) and the Open Project of the Key Laboratory of Computational Physical Sciences (Ministry of Education). Z.X.G. thanks the Fundamental Research Funds for Central Universities (No. xzy012019062) and Open Research Fund of Key Laboratory of Polar Materials and Devices, Ministry of Education. We gratefully acknowledge HZWTECH for providing computation facilities and the computational resources provided by the HPCC platform of Xi'an Jiaotong University.

## AUTHOR CONTRIBUTIONS
Z.X.G. conceived the idea. P.L. performed the theoretical calculation and numerical simulation. P.L. and X.S.Z. prepared the figures. P.L. and Z.X.G. did data analysis and wrote the paper.

## COMPETING INTERESTS
The authors declare no competing interests.

## ADDITIONAL INFORMATION
**Supplementary information** The online version contains supplementary material available at https://doi.org/10.1038/s41524-022-00706-w.

**Correspondence** and requests for materials should be addressed to Zhi-Xin Guo.